\documentclass[prl,twocolumn,showpacs,amsmath,amssymb,superscriptaddress]{revtex4-1}
\usepackage{graphicx,bm,color,mathptmx,hyperref,dcolumn} 
\newcommand{\Tr}{\mathop{\mathrm{Tr}}\nolimits}

\begin{document}

\title{Efficient algorithm for optimizing data pattern tomography}

\author{L.~Motka}
\affiliation{Department of Optics, Palacky University, 
17. listopadu 12, 77146 Olomouc, Czech Republic}
\author{B.~Stoklasa}
\affiliation{Department of Optics, Palacky University, 
17. listopadu 12, 77146 Olomouc, Czech Republic}
\author{J.~Rehacek}
\affiliation{Department of Optics, Palacky University, 
17. listopadu 12, 77146 Olomouc, Czech Republic}
\email{rehacek@optics.upol.cz}
\author{Z.~Hradil}
\affiliation{Department of Optics, Palacky University, 
17. listopadu 12, 77146 Olomouc, Czech Republic}
\author{V.~Karasek}
\affiliation{Department of Optics, Palacky University, 
17. listopadu 12, 77146 Olomouc, Czech Republic}
\author{D.~Mogilevtsev}
\affiliation{Institute of Physics, 
Belarus National Academy of Sciences,
F Skarina Ave 68, Minsk 220072, Belarus}
\author{G.~Harder}
\affiliation{Department of Physics, University of Paderborn, 
Warburger Stra{\ss}e 100, 33098 Paderborn, Germany}
\author{C.~Silberhorn}
\affiliation{Department of Physics, University of Paderborn, 
Warburger Stra{\ss}e 100, 33098 Paderborn, Germany}
\affiliation{Max-Planck-Institut f\"ur die Physik des Lichts,
  G\"{u}nther-Scharowsky-Stra{\ss}e 1, Bau 24, 91058 Erlangen,
  Germany}
\author{L.~L.~S\'{a}nchez-Soto}
\affiliation{Max-Planck-Institut f\"ur die Physik des Lichts,
  G\"{u}nther-Scharowsky-Stra{\ss}e 1, Bau 24, 91058 Erlangen,
  Germany} 
\affiliation{Departamento de \'Optica, Facultad de F\'{\i}sica, 
Universidad Complutense, 28040 Madrid, Spain}

\begin{abstract}
  We give a detailed account of an efficient search algorithm
  for the data pattern tomography proposed by J.~Rehacek,
  D.~Mogilevtsev, and Z.~Hradil [Phys.~Rev.~Lett.~\textbf{105}, 010402
  (2010)], where the quantum state of a system is reconstructed
  without \emph{a priori} knowledge about the measuring setup. The
  method is especially suited for experiments involving complex
  detectors, which are difficult to calibrate and characterize. We
  illustrate the approach with the case study of the homodyne
  detection of a nonclassical photon state.
\end{abstract}


\maketitle

Modern quantum technologies rely on the ability to create, manipulate,
and measure quantum states.  For the successful completion of these
tasks, verification of each step in the experimental procedures is of
utmost importance: quantum tomography has been developed for 
that purpose~\cite{lnp:2004uq,Chuang:2000fk,Andersen:2010uq}.

The main challenge of tomography is simple to state: given a finite
set of identical copies of a system in a state represented by
the density matrix $\varrho$, and an informationally complete
measurement~\cite{Prugovecki:1977fk,*Busch:1989kx,*Ariano:2004kx,
*Flammia:2005fk,*Weigert:2006zr}, the state $\varrho$ must be inferred
from the measured relative frequencies $f_{\ell}$, which sample the
true probabilities $p_{\ell}$ of distinct measurement outcomes. With
these limited resources, the choice of optimal measurements and the
design of efficient reconstruction algorithms turn out to be decisive.

The standard tomographic approach assumes a well-described measurement
apparatus, that is, the responses $\varrho \mapsto \{ p_{\ell} \}$ to all
the states in the search space can be determined. The issue of
 the independent characterization of detectors has recently started to
attract a good deal of attention~\cite{Luis:1999yg,*Fiurasek:2001dn,
*DAriano:2004oe,*Lundeen:2009sf,*Amri:2011fk,*Zhang:2012fu,
 *Brida:2012mz,*Natarajan:2013bh}. Quantum detector tomography employs
the outcome statistics in response to a set of complete certified
input states, at the cost of enlarging the set of unknown parameters
from $d^2$ to $d^4$, in dimension $d$. 

However, as shown in Ref.~\cite{Rehacek:2010fk}, if the measurement
itself is of no interest, the costly detector calibration can be
bypassed by using a direct fitting of data in terms of detector
responses to input probes.  Thus, state estimation is done without any
prior knowledge of the measurement, avoiding unnecessary wasting of
resources on appraising the parameters of the
setup~\cite{Mogilevtsev:2013kl}. In addition, since all the
information used is contained in the data patterns, the method is free
of any assumption that cannot be verified experimentally.  These
substantial advantages have already been experimentally
demonstrated~\cite{Cooper:2013fq}.

The fitting of data patterns requires an optimization process with
additional physical constraints (such as positivity). It is
precisely the goal of this work to present a detailed
implementation of a simple, robust, and efficient algorithm to perform
such a job. This is an essential resource for any potential
practitioner of this promising technique.

We recall that the central idea of the method is the
possibility of expressing an arbitrary quantum signal $\varrho$ as a
mixture,
\begin{equation}
  \label{decomp}
  \varrho = \sum_{\xi}^N x_{\xi} \, \sigma_{\xi}
\end{equation}
of $N$ linearly independent (generally, nonorthogonal) states $\{
\sigma_{\xi} \}$, with positive and negative weights $\{x_{\xi}
\}$. We may look at (\ref{decomp}) as some sort of discrete $P$
representation.

An unknown measurement is mathematically interpreted as a set of
positive operator-valued measures $\{ \Pi_{\ell} \} $, with $\ell = 1,
\ldots, M$ labeling the measurement outcomes~\cite{Helstrom:1976ij}.
The probability for detector outcome $\ell$ given input state
$\sigma_{\xi}$ is given by the Born rule $p_{\ell}^{(\xi)} = \Tr (
\Pi_{\ell} \, \sigma_{\xi})$. In a practical estimation with a finite
number of copies, what we get is a frequency distribution
$f_{\ell}^{(\xi)}$.  By linearity, the response to an unknown signal
$\varrho$ can be written as
\begin{equation}
  \hat{f}_{\ell} = \sum_{\xi}^{N}  x_{\xi} \,  f_{\ell}^{(\xi)} \, . 
\end{equation}
Once the corresponding relative frequencies $f_{\ell}$ are measured,
the coefficients $x_{\xi}$ can be inferred and the signal
reconstructed.

The goodness of the fit can be assessed with a variety of convex
objective functions. In this work, we shall use the square
distance
\begin{equation}
  F  (\{ x_{\xi} \} )  = \sum_\ell^M ( f_{\ell} - \hat{f}_{\ell} )^{2} \, ,
\end{equation}
which provides a robust least-squares fit. Consequently, we have to
minimize the functional $F (\{ x_{\xi} \} )$ subject to  $\varrho
\succeq 0$  and $\Tr (\varrho) = 1$, which ensure that the
reconstructed operator corresponds to a physical state.

The constraint $\Tr (\varrho) = 1$ can be accounted for by noticing
that it implies $x_N=1-\sum_\xi^{N-1}x_\xi$, which leaves us with
$N-1$ independent variables we shall denote, for simplicity, by
$\mathbf{x}= (x_{1}, \ldots, x_{N-1}) \in \mathbb{R}^{N-1}$. To
address the positivity, we employ a continuous function
$c(\mathbf{x})$, such that $c(\mathbf{x}) \ge 0 $ whenever
$\varrho(\mathbf {x}) \succeq 0$ and takes zero value at the boundary
of the convex set of density matrices.
 
Fitting the data patterns thus takes  the simplified form
\begin{eqnarray}
  \label{primal}
  \min_{\mathbf{x}} \quad & &  F (\mathbf{x})  \nonumber \\
  \text{subject to}  \quad & &   c(\mathbf{x}) \ge 0  \, .
\end{eqnarray}
For a convex constraint $c(\mathbf{x} )$, the primal
problem~\eqref{primal} is convex  and strictly feasible. 
The dual problem associated with Eq.~\eqref{primal} can be stated
as~\cite{Forsgren:2002sy,Boyd:2004qd}
\begin{eqnarray}
  \label{dual}
  \max_\lambda \, \min_{\mathbf{x}}  \quad & & 
  \mathcal{L} (\mathbf{x},\lambda)
  \nonumber \\
  \text{subject to} \quad  & & \lambda \ge 0 \, ,
\end{eqnarray}
with $\cal{L}$ being the Lagrangian of~\eqref{primal}
\begin{equation}
  \label{lagrang}
  \mathcal{L}(\mathbf{x},\lambda) = 
  F (\mathbf{x} ) - \lambda c(\mathbf{x} ) \, ,
\end{equation}
and $\lambda$ being a dual variable.  For strictly feasible convex
problems, strong duality holds: the optimal of the
Lagrange dual problem coincides with the minimum for the primal
problem.

The complementary slackness condition ensures that at the local
optimum $x^{\ast}$ of (\ref{dual}) one has $\lambda
c(\mathbf{x}^\ast)=0$, and therefore the Lagrange multiplier
must be zero when the constraint is not active at $x^{\ast}$. 
The complementary is perturbed by introducing  a parameter $\mu$
\begin{equation}
  \label{perturbed}
  \lambda c( \mathbf{x})=\mu \, ,
\end{equation} 
to keep the search direction biased from the boundary. 

The optimality conditions for \eqref{primal} [or, equivalently,
for~\eqref{dual}] are tantamount to including a logarithmic barrier
function~\cite{Forsgren:2002sy}; that is, instead of the constrained
problem (\ref{primal}), one looks at the unconstrained version
\begin{equation}
  \label{barrier}
  \min_{\mathbf{x}} \quad 
  F  ( \mathbf{x}) -\mu \log  c ( \mathbf{x})  \, .
\end{equation}
Given the properties of $c(\mathbf{x})$, the barrier term $ \log
c(\mathbf{x}) $ goes to infinity as the point approaches the boundary
of the feasible region. In this way, it penalizes points close to the
border and thus ensures that one searches for an optimum well inside
the region where the constraint is satisfied.  The barrier
parameter $\mu$ plays the role of a scaling factor: when it becomes
very small the effect of the barrier becomes negligible within the
strictly feasible set and only remains at the border.

By choosing a feasible starting point and gradually decreasing the
height of the barrier, the optimal points of~\eqref{barrier} will
converge to the optimal points of the primal problem \eqref{primal}
from the interior regardless of the purity of the optimal state.

The extremal equations for the dual problem (\ref{dual}) read
\begin{equation}
  \label{extremal}
  g (\mathbf{x} ) -\lambda J (\mathbf{x}) =0 \, ,
\end{equation}
along with the constraint \eqref{perturbed}. Here, $g (\mathbf{x})
=\nabla F (\mathbf{x} )$ and $J (\mathbf{x})=\nabla c(\mathbf{x})$ is
the constraint Jacobian.  There are a variety of numerical methods to
solve (\ref{extremal}), although the Newton search provides a
particularly fast convergence.  The Newton steps $\Delta_{\mathbf{x}}$
and $\Delta_{\lambda}$ of the primal and dual variables, respectively,
obey 
\begin{equation}
  \label{newton}
  \left(
    \begin{array}{cc}
      H & -J^T\\
      \lambda J & c
    \end{array}
  \right)\left(
    \begin{array}{c}
      \Delta_{\mathbf{x}} \\
      \Delta_{\lambda}
    \end{array}
  \right)=\left(
    \begin{array}{c}
      -g+\lambda J^T \\
      \mu-\lambda c
    \end{array}
  \right),
\end{equation}
where $H_{ij}=\partial^2\mathcal{L}(\mathbf{x},\lambda)/\partial
x_i\partial x_j$ is the Hessian matrix of the Lagrangian
\eqref{lagrang}.  To proceed further, we need to specify the function
$c(x)$.  Motivated by the barrier function $\log \det
\varrho(\mathbf{x})^{-1}$, which is strictly convex and analytical on
the feasible space~\cite{Vandenberghe:1996ca}, and has been already
employed in maximum likelihood estimations~\cite{Moroder:2012bs}, we
propose to adopt 
\begin{equation}
  c(\mathbf{x})=
  \left\{
    \begin{array}{ll}
      [ \det \varrho(\mathbf{x}) ]^m \, ,  & \qquad \varrho(\mathbf{x})\succeq
      0 \, ,\\
      & \\
      0,  & \qquad \text{otherwise ,}
    \end{array}
  \right.
\end{equation}
where the parameter $m$ ($0<m<1$) has been inserted to deal with the
numerical issues that arise due to the extremely small values of
$\det\varrho$ near a highly rank-deficient optimum. Setting $m$ to be
the reciprocal of the Hilbert-space dimension $m=1/d$ works well, and
the algorithm is not very sensitive to small changes  in this
suggested value.
 
Using simple matrix  identities, we get
\begin{equation}
  \begin{split}
    J_i (\mathbf{x})&\equiv 
  \frac{\partial c(\mathbf{x})}{\partial x_{i}} =
   m c(\mathbf{x}) \Tr(\Gamma_i),\\
    B_{ij}(\mathbf{x})&\equiv 
   \frac{\partial^2 c(\mathbf{x})}{\partial x_i x_j} = 
   c(\mathbf{x})^{-1} J_i J_j - m c(\mathbf{x})  
\Tr(\Gamma_i\Gamma_j),
  \end{split}
\end{equation}
where we have  denoted $\Gamma_i=\varrho^{-1}(\sigma_i-\sigma_N)$. 

With all these results in mind, we are ready to work out the desired
solution. Our algorithm consists of outer and inner
iterations; the latter solve \eqref{extremal} and \eqref{perturbed}
for a fixed value of $\mu$. This value is gradually decreased to zero
in outer iterations.  In practice, only one inner iteration is done
per outer iteration to increase the rate of convergence.  The
algorithm can be summarized in the following steps:
\begin{description}
\item [\sc{step 1}] Choose $\mu\ge 0$ and $0\le \beta \le 1$. Set
  $x_i=1/N$ and $\lambda=\mu/c( \mathbf{x} )$.

\item [\sc{step 2}] Solve the system \eqref{newton} for the
  primal-dual steps $\Delta_{\mathbf{x}}$ and $\Delta_{\lambda}$.

\item [\sc{step 3}] Set $\mathbf{x}^{\prime}=\mathbf{x}+\alpha
  \Delta_{\mathbf{x}}$ and $\lambda^{\prime}=\lambda+\alpha
  \Delta_{\lambda}$. Start from $\alpha = 1$ and backtrack
  $\alpha$ until $\varrho(x^{\prime})\succeq 0$, $\lambda^{\prime}\ge
  0$, and a sufficient decrease of the residuals of \eqref{extremal}
  is observed.

\item [\sc{step 4}]  Decrease the barrier parameter $\mu=\beta
  \lambda^{\prime} c(\mathbf{x}^{\prime})$ and  update the variables
  $\mathbf{x}= \mathbf{x}^{\prime}$ and $\lambda=\lambda^{\prime}$.

\item [\sc{step 5}] Repeat from {\sc{step 2}} until convergence.
\end{description}

Fine tuning of the algorithm can be achieved by altering the initial
values of $\mu$ and $\beta$. Larger values tend to slow down the
convergence, but improve stability.  Typically, a few tens of
iterations are required to solve a moderately sized problem (say,
$d\approx 7$, $M\approx 80$, and $N\approx 100$).

The complexity of a single iteration depends on the parameters $d$,
$M$, and $N$. Since there are at most $M-1$ linearly independent
normalized patterns of size $M$, we can always set $N<M$.  Three
exclusive cases of interest can be identified: 
\begin{enumerate}
\item Oversampled measurements: $M>\max(d^3,N d^2, N^2)$.  
 Setting up  $\hat{f}_{\ell} $ and $\nabla F  (\mathbf{x})$ dominates
 with cost $O(N M)$.

\item  Informationally incomplete measurements:  $N<d^2$.  
Setting up  the constraint Hessian dominates with costs $O(N
d^3)$ and $O(N^2 d^2)$  to generate all $\Gamma_i$'s and carry out 
the pairwise inner products of $\Gamma_i$ and $\Gamma_j$.

\item Informationally complete measurements:  $N\ge d^2$.  
Solving the system \eqref{newton} dominates with cost
$O(N^3)$.
\end{enumerate}
The complexity can be decreased by adopting a quasi-Newton
approach~\cite{Fletcher:1987bx} with a Broyden-Fletcher-Goldfarb-Shanno
(BFGS) update of the Hessian matrix [case (2)] or a Hessian matrix
inverse [case (3)] at the cost of slowing down the 
convergence~\cite{Armand:2001ys}.

\begin{figure}
  \centerline{
    \includegraphics[width=0.7\columnwidth]{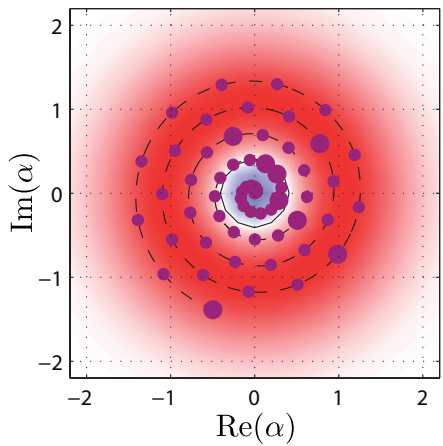}}
\caption{(Color online) Amplitudes of coherent probes used to fit 
the data  pattern  in  homodyne tomography.  Probes 13, 15,
  16, 25, 30, 40,   50, and 60 are marked with larger symbols. The
  density plot of the  true $W(\alpha)$ is shown in the background:
  white represent  zero of  $W(\alpha)$, while the external part
  (red) and the internal one (blue) are the zones where $W(\alpha)$ is
  positive and negative, respectively.}
  \label{fig:spiral}
\end{figure}

To illustrate the utility of the proposed algorithm we examine the
case of the homodyne measurement of a nonclassical photon state.  We
are then concerned with rotated-quadrature measurements $x(\theta)= x
\cos \theta+ p \sin \theta$, where $x$ and $p$ are the basic optical
position and momentum observables and $\theta$ is the phase of the
local oscillator.  With a realistic detector efficiency of
$\eta=80~\%$, the measurement consists of eigenvectors of $x(\theta)$
quadratures convolved with the vacuum. Explicit formulas for the
measurement operators in the computational Fock basis can be found,
e.g., in Ref.~\cite{Lvovsky:2004oq}.  We discretize the measurement
using six equidistant phases and 61 quadrature value bins in
the interval $x \in [-6, 6]$. Each of these six quadratures is
measured $200 000$~times for each different state, with the data being
drawn from the multinomial distribution describing the measurement
statistics.

\begin{figure}
  \centerline{\includegraphics[width=0.66\columnwidth]{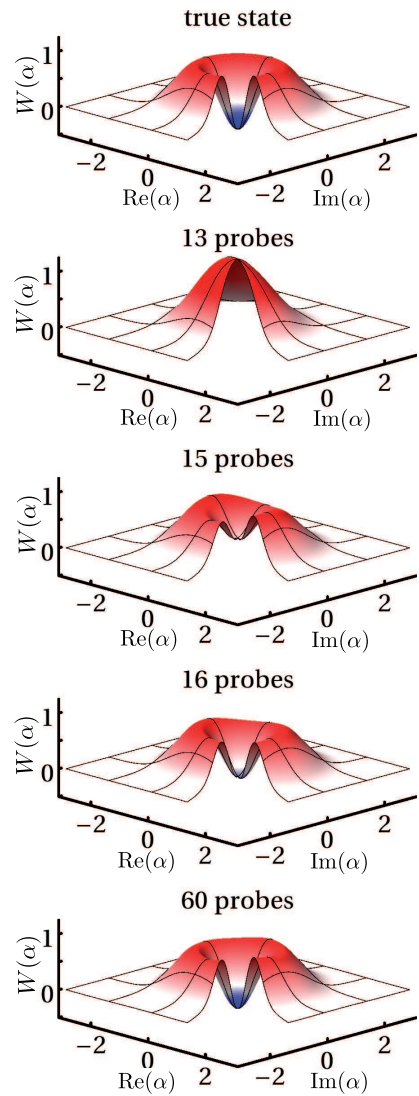}}
  \caption{(Color online) True vs. reconstructed Wigner functions for
    different  number of coherent probes. All the reconstructions have
    been  performed in an eight-dimensional Fock subspace.} 
  \label{fig:wigner}
\end{figure}

As a signal state, we have simulated an incoherent mixture
$\varrho_{\mathrm{true}}=0.4 |0\rangle\langle 0|+ 0.6 |1\rangle\langle
1|$ of vacuum and a single-photon state, which can be prepared in
parametric downconversion~\cite{Laiho:2010kl}. The pronounced
negativity of the corresponding Wigner function at the origin is a
nonclassicality witness and will be a test for our scheme.  As with
every genuine quantum feature, it is rather sensitive to tomography
imperfections.

\begin{figure}
  \centerline{\includegraphics[width=0.85\columnwidth]{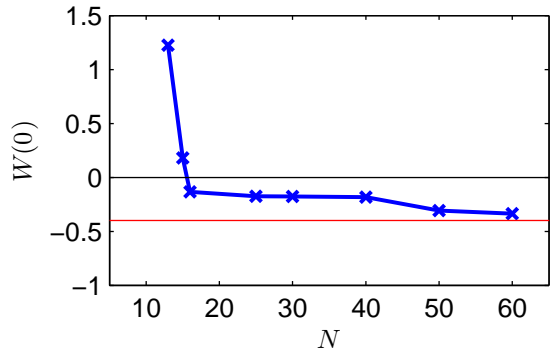}}
\caption{(Color online) Reconstructed Wigner functions evaluated at
  the origin, $W(0)$,  with 13, 15, 16, 25, 30, 40, 50, and 60
  coherent probes, as   in Fig.~\ref{fig:spiral}.  The
  quantum/classical border, $W(0) =0$, and   the true negative value
  $W(0)=-0.4$  are indicated by horizontal lines.}
  \label{fig:negativity}
\end{figure}

In our simulation, this state is measured together with a set of known
coherent probe states $\sigma_i=|\alpha_i\rangle\langle \alpha_i|$,
which are robust and easy to generate on demand.  The coherent
amplitudes are sampled from a spiral pattern unwinding from the
origin, as sketched in Fig.~\ref{fig:spiral}. The resulting samples
are equidistant in radius and angle, but other choices, such as a
rectangular grid, would work as well.

Without \emph{a priori} knowledge of the true state, one might think
of sampling the phase space starting from the origin and gradually
increasing the size of the probe set until no significant updates of
the reconstruction are observed. This strategy is illustrated in
Fig.~\ref{fig:wigner}, which shows the reconstructed Wigner function
for different numbers of coherent probes. Notice that 13 probes yield
a classical state whose Wigner function peaks near the origin.  The
central dip develops with 15 probes, and just 16 probes are enough to
observe negativities.  Finally, with 60 probes the reconstruction
becomes nearly perfect, with some residual errors due to unavoidable
statistical noise.  The reconstructed $W(\alpha)$ becomes smoother and
circularly symmetrical with larger probe sets, increasing thus the
overall fidelity of the protocol [the fidelity $\mathcal{F}=
\Tr(\sqrt{\sqrt{\varrho_{\mathrm{true}}} \varrho
  \sqrt{\varrho_{\mathrm{true}}}})$ of the reconstruction with 60
probes is $99.2~\%$].

In Fig.~\ref{fig:negativity} we plot the reconstructed value of $W(0)$
as a function of the probe set size $N$.  It is intriguing to observe
that the abrupt drop of $W(0)$ for $N \sim 16$ arises when the probe
amplitudes reach the edge of the negative region of the true Wigner
function, as can be seen in Fig.~\ref{fig:spiral}. Furthermore, the
drop of $W(0)$ between $N=40$ and $N=50$ seems to happen at the point
where the coherent probes pass the maximum of $W$ and start to feel
the region in which the true Wigner function decays to zero.

\begin{figure}
  \centerline{
    \includegraphics[width=0.80\columnwidth]{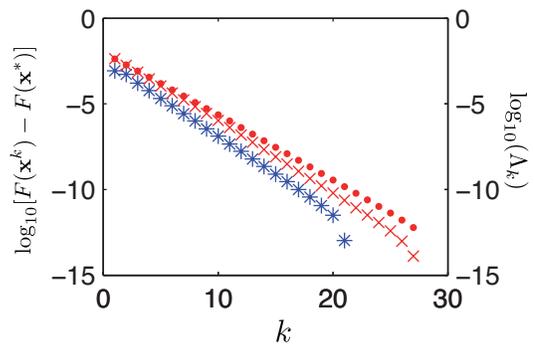}}
  \caption{(Color online) Convergence of the algorithm with 15 (blue
    stars) and 60  probes (red crosses). $F(\mathbf{x}^{\ast})$ is
    the exact value at  the optimal point, while $F(\mathbf{x}^{k})$
    is the calculated  value after the $k$th iteration. For the later
    case, we have also   included $\log ( \Lambda^{k} )$ (in the right
    vertical axis),  where $\Lambda^{k}$ denotes the  minimal
    eigenvalue of $\rho(  \mathbf{x}^{k})$. Observe how the state
    converges towards  the  boundary of the space of density
    matrices. The parameters are  $\mu=0.01$, $\beta=0.1$, $m=1/6$
    and decimal  logarithm is used everywhere.}
  \label{fig:converg}
\end{figure}

In Fig.~\ref{fig:converg} we depict the convergence of the
algorithm for the simulated data in Fig.~\ref{fig:wigner}.  A slight
increase in the number of iterations with the problem size is
observed, as might be expected.
  
\begin{figure}[b]
  \centerline{
    \includegraphics[width=0.8\columnwidth]{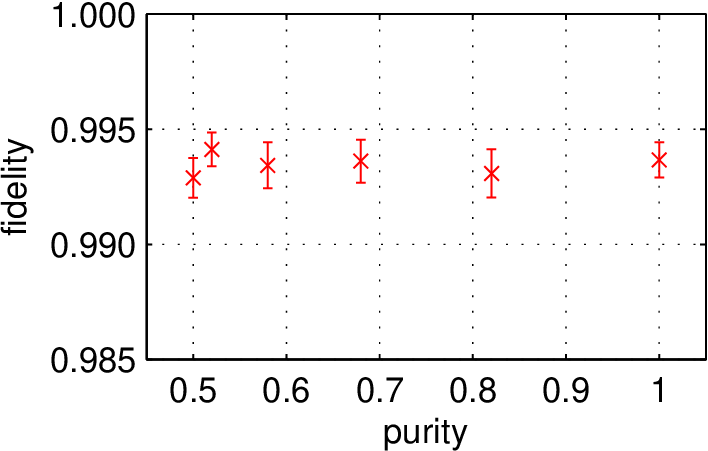}}
  \caption{(Color online)  Mean fidelities of data pattern
    reconstruction with true     states $\rho_{\text{true}}= 
  0.5 |0\rangle\langle  0|+  0.5|1\rangle\langle 1|+
  \gamma |0\rangle\langle 1|+ \gamma|1\rangle\langle 0|$,  
  $\gamma\in [0,0.5]$ of different  purities measured by $\Tr (
  \rho_{\text{true}}^2)$.  The averaging   was done over 50 runs of
  simulated homodyne detection  with 60 probes. Standard deviations
  of  those 50 fidelities are  also shown.}
  \label{fig:purity}
\end{figure}

Last, in Fig.~\ref{fig:purity} we present typical fidelities of
data pattern tomography for states of varying purity.  The variations
of the fidelity observed are not statistically significant.

We stress that our knowledge about the measurement was used solely for
generating data. The pattern tomography itself was based on the signal
data, probe data and the representation of probes in the computational
basis.  In this way, the search space ---the field of view of
tomography--- was defined uniquely by the measured objects, avoiding
the problematic \textit{ad hoc} Hilbert space truncation of the
standard methods~\cite{Mogilevtsev:2013hb}.

In summary, we have re-elaborated on the data pattern approach to
quantum tomography. The most relevant feature of the approach is the
ability to perform an efficient reconstruction without ever knowing
the exact properties of the measurement setup. The knowledge required
for the precise estimation of a particular signal state can be
obtained a posteriori, after the measurement on the signal state. One
can also decide which additional probes might be helpful in further
improving the reconstruction. This is a significant advantage for
experimentalists, since calibrating the measurement setups for
such weak signals can be a rather challenging task.

This work was supported by the European Social Fund and the State
Budget of the Czech Republic POSTUP II (Grant CZ.1.07/2.3.00/30.0041)
and MCIN (Grant CZ.1.07/2.3.00/20.0060), the IGA Project of the
Palack\'y University (Grant PRF 2013 019), the Spanish MINECO (Grant
FIS2011-26786), and the External Fellowship Program of the Russian
Quantum Center at Skolkovo. We acknowledge illuminating discussions
with A. Felipe.


%

\end{document}